\documentclass[11pt,a4paper]{article}
\usepackage{graphics}{}                 
\pdfoutput=1
\usepackage[pdftex]{}

\usepackage{amsmath,amssymb,amsfonts} 
\usepackage{color}                    
\usepackage{hyperref}                 

\begin{document}

\vspace{1.8cm}
\begin{center}
{\Large Ion Program of NA61/SHINE at the CERN SPS}
\end{center}

\vspace*{0.8cm}
\begin{center}
{Marek Gazdzicki\footnote{e-mail address: marek.gazdzicki@cern.ch}
 for the NA61/SHINE Collaboration \\}
\end{center}

\vspace*{0.2cm}
\begin{center}
{\it Institut f\"ur Kernphysik, University of Frankfurt, Frankfurt, Germany\\}
{\it Jan Kochanowski University, Kielce, Poland\\}
\end{center}

\vspace*{0.8cm}
\begin{abstract}
The NA61/SHINE experiment at the CERN SPS aims to discover the critical point
of strongly interacting matter and study properties 
of the onset of deconfinement. These goals will be reached
by measurements of hadron production properties in nucleus-nucleus,
proton-proton and proton-lead interactions as a function of collision
energy and size of the colliding nuclei. 
Furthermore, NA61/SHINE will perform
numerous precision measurements needed  for neutrino (T2K) and
cosmic-ray (Pierre Auger Observatory and KASCADE) experiments.
This paper summarizes physics arguments for the NA61/SHINE ion program
and presents the status and plans of the experiment.
\end{abstract}

\thispagestyle{empty}

\newpage

{\small
\begin{center}
\vspace*{-1.5cm}
The NA61/SHINE Collaboration
\end{center}
\noindent
N.~Abgrall${}^{22}$,
A.~Aduszkiewicz${}^{23}$,
B.~Andrieu${}^{11}$,
T.~Anticic${}^{13}$,
N.~Antoniou${}^{18}$,
A.~G.~Asryan${}^{15}$,
B.~Baatar${}^{9}$,
A.~Blondel${}^{22}$,
J.~Blumer${}^{5}$,
L.~Boldizsar${}^{10}$
% ...

\vspace*{0.1cm}
\noindent
${}^{ 1}$Cape Town University, Cape Town, South Africa \\
${}^{ 2}$ETH, Zurich, Switzerland \\
${}^{ 3}$Fachhochschule Frankfurt, Frankfurt, Germany \\
${}^{ 4}$Faculty of Physics, University of Sofia, Sofia, Bulgaria \\
${}^{ 5}$Forschungszentrum Karlsruhe, Karlsruhe, Germany \\
${}^{ 6}$Institute for Nuclear Research, Moscow, Russia \\
${}^{ 7}$Institute for Particle and Nuclear Studies, KEK, Tsukuba,  Japan \\
${}^{ 8}$Jagiellonian University, Cracow, Poland  \\
${}^{ 9}$Joint Institute for Nuclear Research, Dubna, Russia \\
${}^{10}$KFKI Research Institute for Particle and Nuclear Physics, Budapest, Hungary \\
${}^{11}$LPNHE, University of Paris VI and VII, Paris, France \\
${}^{12}$Pusan National University, Pusan, Republic of Korea \\
${}^{13}$Rudjer Boskovic Institute, Zagreb, Croatia \\
${}^{14}$Soltan Institute for Nuclear Studies, Warsaw, Poland \\
${}^{15}$St. Petersburg State University, St. Petersburg, Russia \\
${}^{16}$State University of New York, Stony Brook, USA \\
${}^{17}$Jan Kochanowski University in  Kielce, Poland \\
${}^{18}$University of Athens, Athens, Greece \\
${}^{19}$University of Bergen, Bergen, Norway \\
${}^{20}$University of Bern, Bern, Switzerland \\
${}^{21}$University of Frankfurt, Frankfurt, Germany \\
${}^{22}$University of Geneva, Geneva, Switzerland \\
${}^{23}$University of Warsaw, Warsaw, Poland \\
${}^{24}$Warsaw University of Technology, Warsaw, Poland  \\
}

\newpage

\section{Introduction}

The objective of the NA61/SHINE ion program is to search for the critical
point of strongly interacting matter and to study  the
properties of the onset of deconfinement~\cite{proposal}.
This goal will be pursued by an experimental investigation
of nucleus-nucleus, proton-proton and proton-lead integrations 
at at the CERN SPS.
NA61/SHINE intends to  carry out -- for the first time in the history
of nucleus-nucleus collisions -- a comprehensive scan in two
dimensional parameter space: the size of the colliding nuclei versus
the interaction energy~\cite{Gazdzicki:2006fy}.
The project has the potential for an important discovery --
the experimental observation of the critical point of
strongly interacting matter. On top of that it guarantees precision
measurements crucial for the understanding of the properties
of the onset of deconfinement.
The synergy of the  NA61/SHINE physics program~\cite{proposal}
on physics of strongly interacting matter with cosmic-ray and neutrino
programs,
as well as the use of the existing accelerator chain and detectors
offers the unique opportunity to reach the  physics goals
in an efficient and cost effective way.

NA61/SHINE
plans to  perform the necessary
measurements
using the upgraded NA49 apparatus~\cite{na49-nim}.
The most essential upgrades are
the increase of data taking and  event rate by a factor of 10 and
the construction of a projectile spectator detector
which will improve the accuracy of determination of the
number of projectile spectators by a factor of about 20.

The NA49 apparatus at the  CERN~SPS served during the last 10 years
as a very reliable, large acceptance hadron spectrometer and
delivered experimental data of high precision over the full
range of SPS beams (from proton to lead)~\cite{na49_beam,Alt:2005zq}
and energies (from
20$A$~GeV to 158$A$~GeV)~\cite{afanasiev:2002mx,Gazdzicki:2004ef}.
Among the most important results from this study is the
observation~\cite{afanasiev:2002mx,Gazdzicki:2004ef}
of narrow structures in the energy dependence of hadron production
in central Pb+Pb collisions.
These structures are found at the low CERN SPS energies
(30$A$--80$A$~GeV) and they are consistent with
the predictions~\cite{Gazdzicki:1998vd}
for the onset of the deconfinement phase transition.
The questions raised by this observation serve as an important  motivation
for new measurements with nuclear beams proposed in the SPS energy range
at the CERN~SPS and
also envisaged at BNL~RHIC~\cite{rhic_low},
FAIR~SIS-300~\cite{cbm} as well as
NICA~MPD~\cite{mpd}.
These programs are to a large extent complementary.
However, specific questions addressed in the NA61/SHINE proposal
can be best studied by NA61/SHINE at the CERN SPS.

\section{Detector, upgrades, performance}

\begin{figure}
\begin{center}
{\resizebox{14 cm}{!}{ \includegraphics{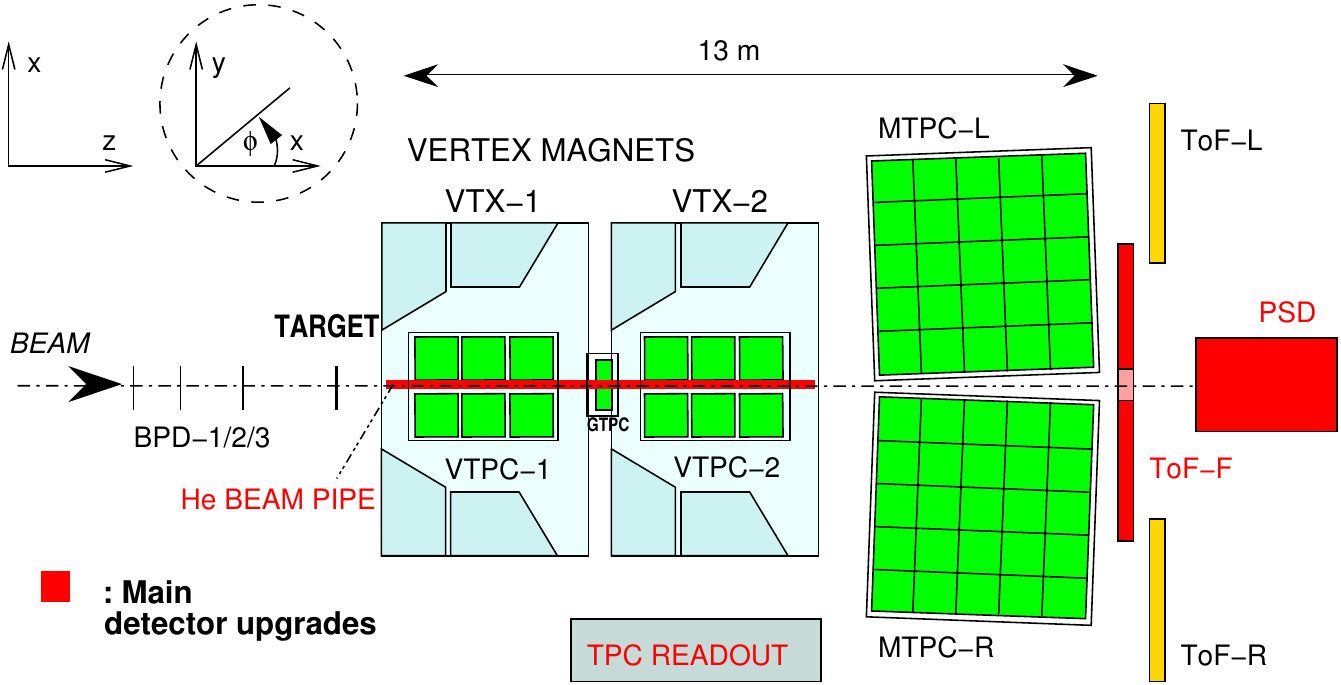} } }
\caption{The layout of the NA61/SHINE setup 
(top view, not to scale) with the basic upgrades
indicated in red.}

\end{center}
\label{na61-setup}
\end{figure}

The NA61/SHINE experiment uses a large acceptance hadron spectrometer
at the CERN-SPS
for the study of the hadronic final states produced in interactions of
various beam particles ($\pi$, p, C, S and In)
with a variety of fixed targets at the SPS energies.
The layout of the NA61/SHINE set-up is shown in Fig.~\ref{na61-setup}.
The main components of the current detector
were constructed and used by the NA49 experiment~\cite{na49-nim}.
The main tracking devices are four large volume
Time Projection Chambers (TPCs in Fig.~\ref{na61-setup}),
which are capable of detecting up to 70\% of all charged particles
created in the studied reactions.
Two of them, the vertex TPCs (VTPC-1 and VTPC-2), are
located in the
magnetic field of two super-conducting dipole magnets (maximum bending
power of 9~Tm)
and two others (MTPC-L and MTPC-R) are positioned
downstream of the magnets symmetrically with respect to the beam line.
One additional small TPC, the so-called gap TPC (GTPC), is installed on
the beam axis between the vertex TPCs.
The setup is supplemented by time of flight detector arrays
two of which (ToF-L/R) were inherited from NA49 and can provide
a time measurement resolution of $\sigma_{tof} \approx 60$~ps.

In a 2007 pilot run data for the T2K neutrino experiment 
on p+C interactions at 31~GeV/c were recorded~\cite{na61-calib}.
For this run a new forward time of flight detector (ToF-F) was
constructed in order to extend in these low-multiplicity events the acceptance
for pion and kaon identification to low momenta ($1 < p < 4$~GeV/c).
The ToF-F wall is installed  downstream of 
 TPC-L and MTPC-R (see Fig.~\ref{na61-setup}),
closing the gap between the ToF-R and ToF-L walls.
Furthermore, one super-module of the Projectile Spectator Detector (PSD)
was installed downstream of ToF-F and tested~\cite{PSD}.
The required energy resolution 
$\sigma(E)/E \approx 50\%/\sqrt{E/(1~{\rm GeV})}$
was reached.

A major step forward in the detector performance, 
i.e. the TPC readout and DAQ upgrade,
was achieved for the 2008 run.
During the run the upgrade was tested.
It results in an increase of the data rate by a factor of about 10
compared to the NA49 rate.
The 2008 run was cut short due to the LHC incident. 

Compatibility with the I-LHC schedule requires the use of
{\it secondary} ion beams for the NA61/SHINE ion program.
Feasibility studies are in progress and first results 
indicate that the secondary ion beam properties are sufficient
to reach the physics goals of NA61/SHINE.

\section{Data Taking Plan}

The NA61/SHINE data taking plan is presented in Table~\ref{beam2}.
The data taking program starts with $p$ and $\pi^-$ beams for the T2K
and cosmic ray experiments.
Three energy scans with the secondary ion beams are planned with nuclei of
mass numbers of $A \approx$ 10, 30 and 100.
In addition, energy scans of p+p and p+Pb collisions will be performed.
They will include high statistics runs at 158 GeV dedicated for a study 
of hadron production at large transverse momenta.

\begin{figure}[!t]
\begin{center}
{\resizebox{12 cm}{!} { \includegraphics{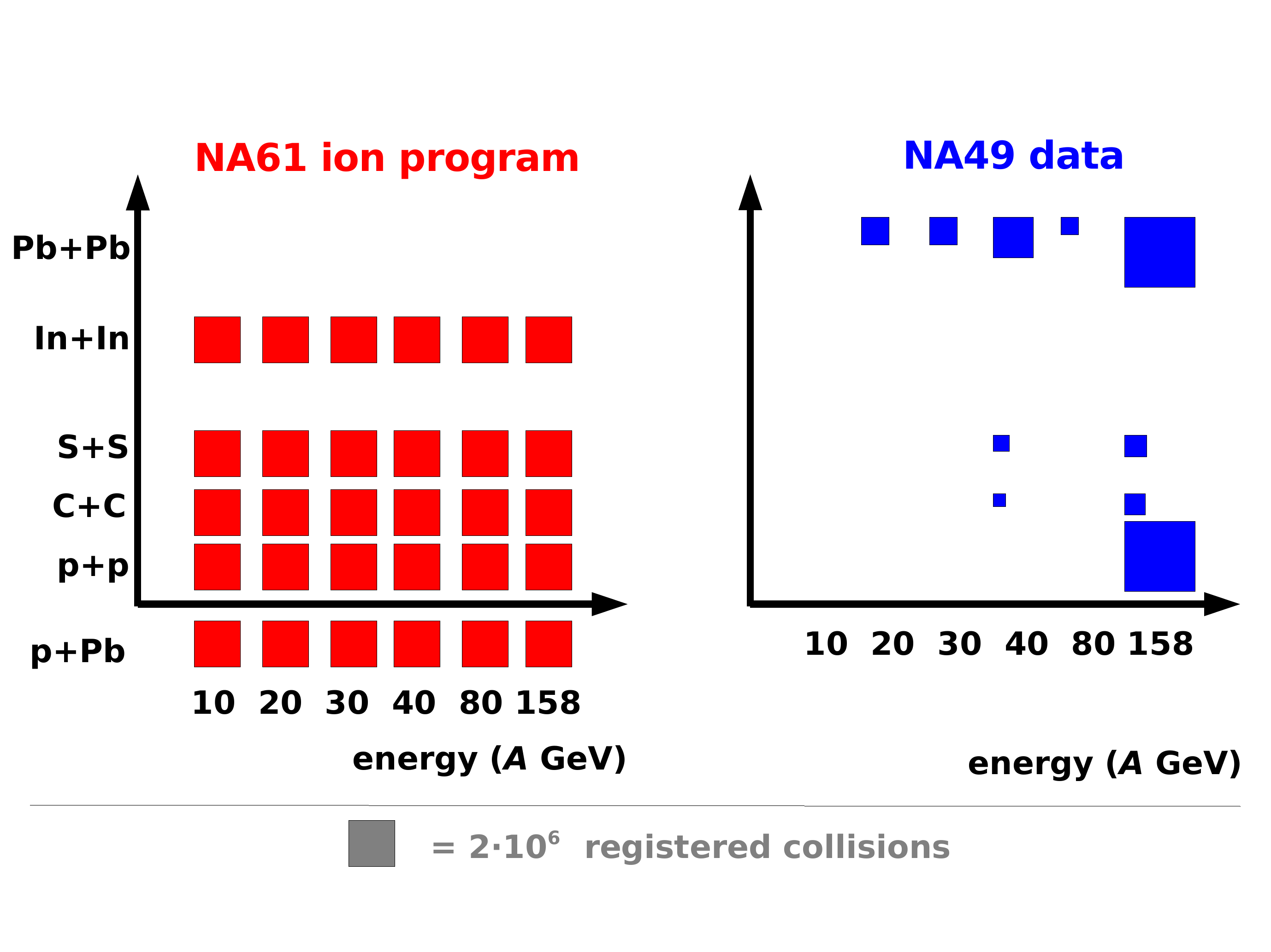} } }
\caption{
The data sets planned to be recorded by NA61/SHINE (left) with the ion program
and those recorded by NA49 (right). The area of the boxes is proportional
to the number of registered central collisions, which for NA61/SHINE will be
$2 \cdot 10^6$ per reaction. 
}
\end{center}
\label{runs}
\end{figure}

{\bf
\begin{table}[!th]
\begin{center}
\begin{tabular}
{ l r r r r r }
\hline
  Beam+Target &  Energy    &  Year  & Days &  Physics & Status\\
       & ($A$ GeV)  &        &      &       &   \\
\hline
\hline
\bf p+C(T2K)  & \bf 30   & \bf 2007  & \bf 30
& \bf  T2K, C-R & \it performed  \\
\hline
\hline
\bf p+C(T2K)  & \bf 30   & \bf 2009  & \bf 21
& \bf  T2K, C-R & \it recommended  \\
\bf $\pi^-$+C & \bf 158, 350   & \bf 2009  & \bf 14
& \bf  C-R & \it recommended \\
\bf p+p  &\bf  10, 20, 30, 40, 80, 158  &\bf  2009  &\bf  49
& \bf CP\&OD & \it recommended   \\
\hline
\hline
\bf p+p  & \bf 158          & \bf 2010  & \bf 77
&\bf  High p$_T$ & \it recommended   \\
\hline
\hline
\bf $\approx$(30+30)  & \bf 10, 20, 30, 40, 80, 158  & \bf 2011  &\bf 42
&\bf  CP\&OD  & \it recommended  \\
\bf p+Pb  & \bf 158          & \bf 2011  & \bf 42
&\bf  High p$_T$ & \it recommended   \\
\hline
\hline
\bf $\approx$(10+10)   & \bf 10, 20, 30, 40, 80, 158  &\bf  2012  &\bf 42
&\bf  CP\&OD  & \it to be discussed \\
\bf p+p  &\bf  10, 20, 30, 40, 80, 158  &\bf  2012  &\bf 42
& \bf  CP\&OD  & \it recommended \\
\hline
\hline
\bf $\approx$(100+100)  & \bf 10, 20, 30, 40, 80, 158  &\bf 2013  &\bf 42
&\bf  CP\&OD  & \it to be discussed  \\
\hline
\hline
\end{tabular}
\end{center}
\caption[dummy]{
The NA61/SHINE  data taking plan. 
The runs with secondary ion beams are planned for 2011, 2012 and
2013. In these runs the nuclear mass number of the selected ions
will be $A \approx 30$,  $A \approx 10$ and  $A \approx 100$, respectively.
The following abbreviations are used for the
physics goals of the data taking: CP - Critical Point, OD - Onset of
Deconfinement, C-R - Cosmic Rays.
}
\label{beam2}
\end{table}
}
The data sets planned to be recoded by NA61/SHINE for the ion program
and those recorded by NA49 are compared in Fig.~\ref{runs}.

\section{Experimental landscape}

\begin{figure}[!t]
\begin{center}
{\resizebox{12 cm}{!} { \includegraphics{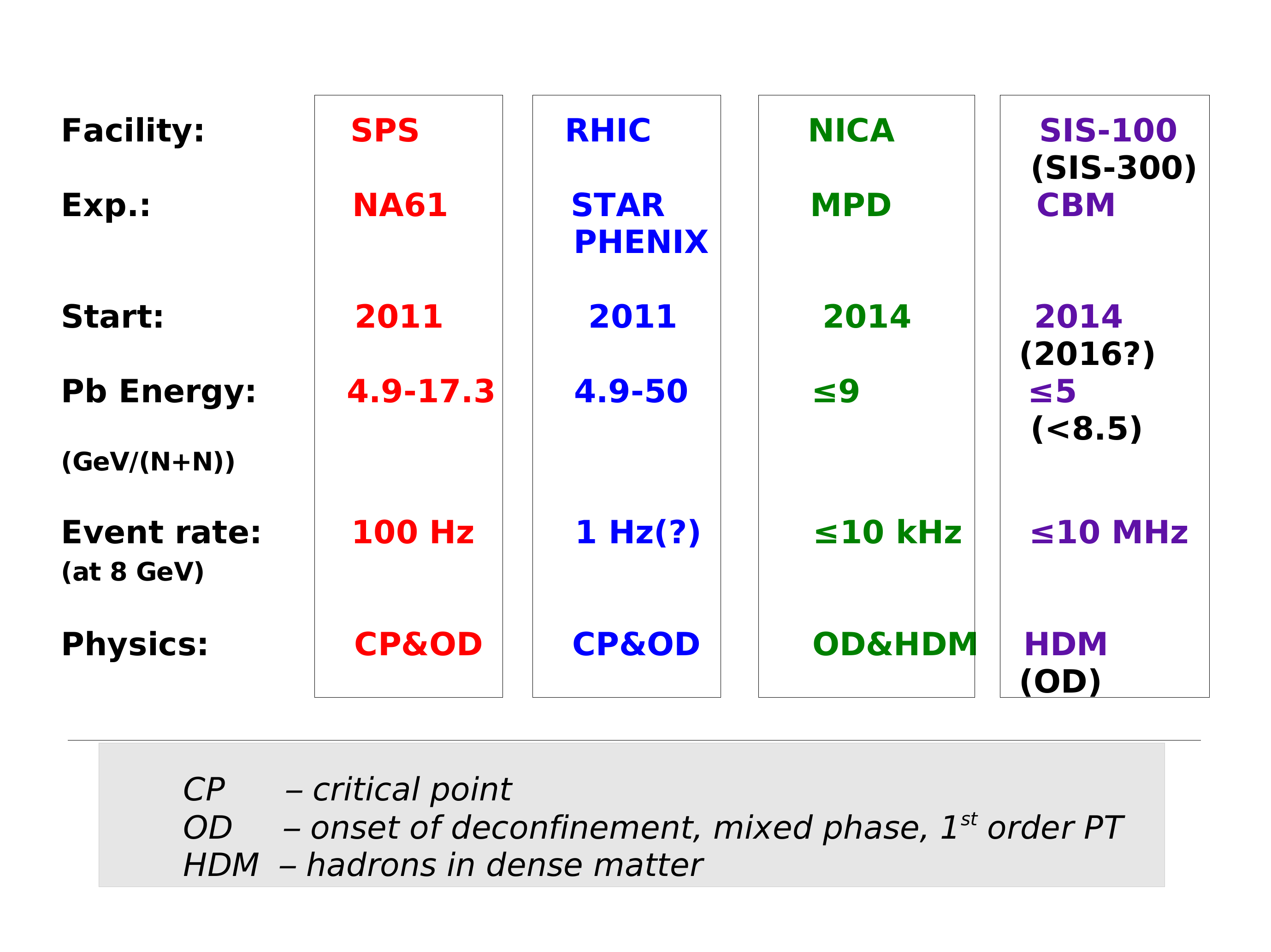} } }
\caption{
The main parameters of the proposed experimental programs for the study of
nucleus-nucleus collisions at SPS, RHIC, NICA and SIS-300. 
}
\end{center}
\label{comp}
\end{figure}

The exciting and rich physics which can be studied in
nucleus-nucleus collisions at the CERN SPS energies
motivates groups from BNL, JINR and FAIR to propose
experimental studies which will complement the CERN SPS
program.
Two fixed target programs (CERN SPS~\cite{proposal} and
FAIR SIS-300~\cite{cbm}) and two
programs with ion colliders (BNL RHIC~\cite{rhic_low}
and JINR NICA~\cite{mpd}) are foreseen.
The basic parameters of the future programs are summarized in
Fig.~\ref{comp}.
The SPS and RHIC energy range covers energies from below
up to well above the energy of the
onset of deconfinement ($\approx$30$A$~GeV in
the fixed target mode). Thus these machines are well suited for the
study of the properties of the onset of deconfinement and the search for the
critical point.

\begin{figure}[!t]
\begin{center}
{\resizebox{15 cm}{!} { \includegraphics{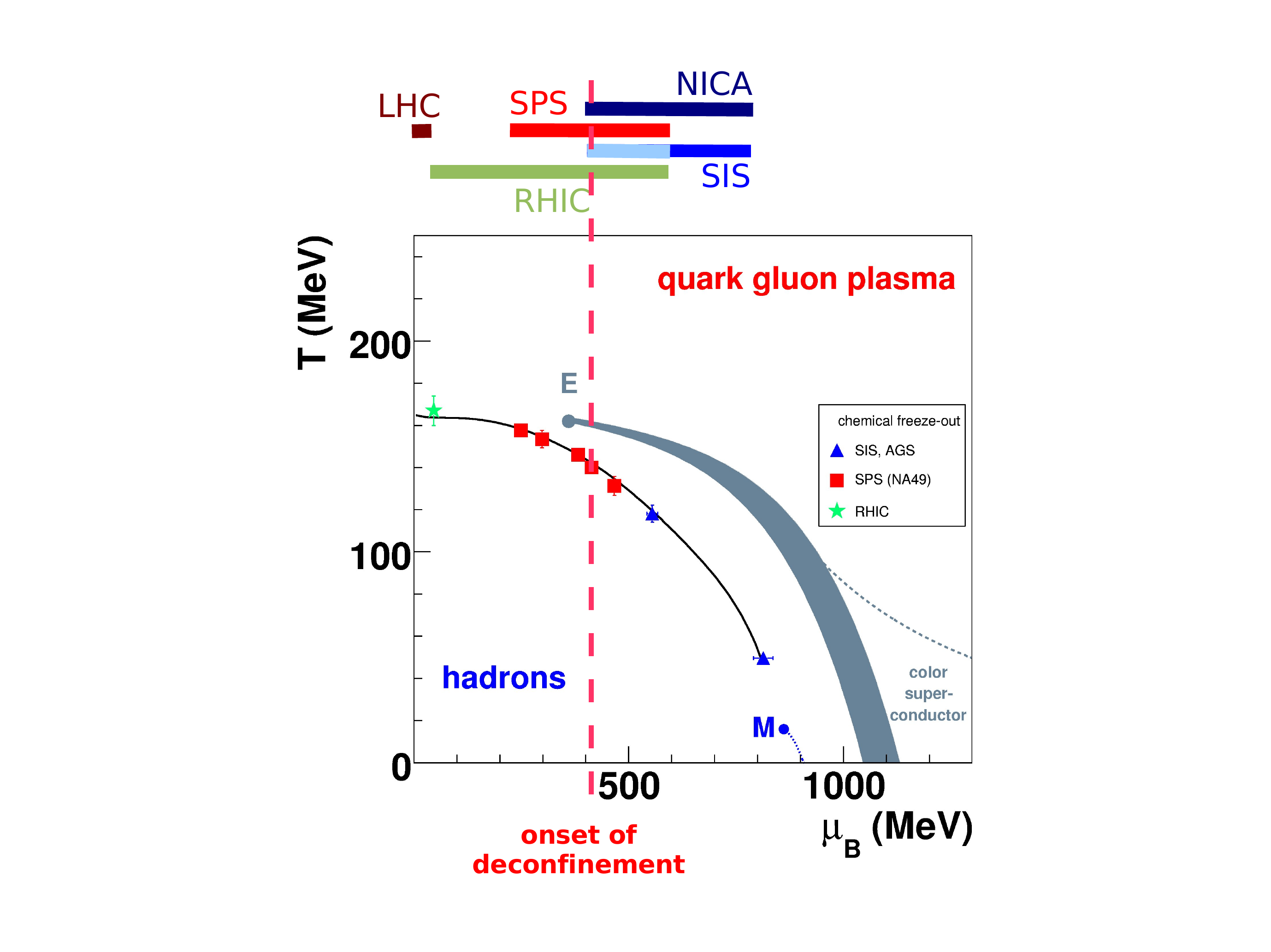} }}
\caption{
The phase diagram of strongly interacting matter with indicated
chemical freeze-out points of central Pb+Pb (Au+Au) collisions
at different energies, and baryon-chemical potential ranges
covered by the future programs.
}

\end{center}
\label{phase_acc}
\end{figure}

The top energies of NICA and SIS-300 are just above the energy of the
onset of deconfinement. The physics at these machines will thus focus
on the study of the properties of dense confined matter
close to the transition to the QGP.
This is  illustrated in Fig.~\ref{phase_acc}
which shows the  coverage of the new programs in the
baryon chemical potential together with the existing data and physics
benchmarks in the $T - \mu_B$ plane.

The advantages of the NA61/SHINE ion program over 
the RHIC energy scan program are:
\begin{itemize}
\item
measurements of identified hadron spectra in a broad
rapidity range, which in particular, allow to obtain the mean hadron multiplicities
in full phase-space,
\item
measurements of the total number of the projectile spectator
nucleons including free nucleons and nucleons in nuclear
fragments,
\item
high  event rate  in the full SPS energy range
including the lowest energies,
\item
high flexibility in selecting the nuclear mass number (thanks to
the secondary ion beam option) and energy (thanks to the SPS features)
of the projectile ions.
\end{itemize} 

The importance of the physics questions addressed by the SPS and RHIC
programs requires two independent, partly complementary measurements.
Thus, both planned experimental programs should be performed.

\vspace{0.5cm}

{\bf Acknowledgments:} 
This work was supported by
the Virtual Institute VI-146 of Helmholtz Gemeinschaft, Germany,
Korea Research Foundation (KRF-2008-313-C00200),
the Hungarian Scientific Research Fund (OTKA 68506),
the Polish Ministry of Science and Higher Education (N N202 3956 33),
the Federal Agency of Education of the Ministry of Education and Science
of the Russian Federation (grant RNP 2.2.2.2.1547) and
the Russian Foundation for Basic Research (grant 08-02-00018),
the Ministry of Education, Culture, Sports, Science and Technology,
Japan, Grant-in-Aid for Scientific Research (18071005, 19034011,
19740162),
Swiss Nationalfonds Foundation 200020-117913/1
and ETH Reseach Grant TH-01 07-3.


\end{document}